\documentclass[notitlepage, superscriptaddress
,reprint 
]{revtex4-2} 

\usepackage{color}
\usepackage{graphicx}
\usepackage{physics}
\usepackage{bbm}
\usepackage{bbold}
\usepackage[normalem]{ulem}
\usepackage{amsmath}
\usepackage{bm}
\usepackage{amssymb}
\usepackage{xcolor}
\usepackage{tikz}
\usetikzlibrary{quantikz2}
\usepackage{diagbox}
\usepackage{booktabs}

\usepackage{hyperref}
\hypersetup{
    colorlinks=true,
    linkcolor=blue, 
    urlcolor=cyan
}

\usepackage{lineno}

\usepackage[caption=false]{subfig}
\usepackage{ragged2e}    
\usepackage{etoolbox}    

\makeatletter
\patchcmd{\@makecaption}
  {\fnum@figure\acap@delim\space\ignorespaces\@tempd}
  {\fnum@figure\acap@delim\space\ignorespaces\justifying\@tempd}
  {}{}
\makeatother

\begin{document}

\title{Probing chiral topological order of the Kitaev honeycomb on a trapped ion processor}

\author{Ammar Ali}
\affiliation{Department of Physics and Astronomy, Purdue University, West Lafayette, IN, 47906, USA}

\author{Joe Gibbs}
\affiliation{School of Mathematics and Physics, University of Surrey, Guildford, GU2 7XH, UK}
\affiliation{AWE, Aldermaston, Reading, RG7 4PR, UK}

\author{Keerthi Kumaran}
\affiliation{Department of Physics and Astronomy, Purdue University, West Lafayette, IN, 47906, USA}

\author{Varadharajan Muruganandam}
\affiliation{Department of Physics and Astronomy, Purdue University, West Lafayette, IN, 47906, USA}

\author{Bo~Xiao}
\affiliation{Materials Science and Technology Division, Oak Ridge National Laboratory, Oak Ridge, TN 37831, USA}

\author{Paul Kairys}
\affiliation{Quantum Information Science Section, Oak Ridge National Laboratory, Oak Ridge TN 37831, USA}

\author{G\'abor B. Hal\'asz}
\affiliation{Materials Science and Technology Division, Oak Ridge National Laboratory, Oak Ridge, TN 37831, USA}

\author{Arnab Banerjee}
\email[]{Arnabb@purdue.edu}
\affiliation{Department of Physics and Astronomy, Purdue University, West Lafayette, IN, 47906, USA}

\author{Phillip C. Lotshaw}
\email[]{Lotshawpc@ornl.gov}
\thanks{\\ This manuscript has been authored by UT-Battelle, LLC, under Contract No. DE-AC0500OR22725 with the U.S. Department of Energy. The United States Government retains and the publisher, by accepting the article for publication, acknowledges that the United States Government retains a non-exclusive, paid-up, irrevocable, world-wide license to publish or reproduce the published form of this manuscript, or allow others to do so, for the United States Government purposes. The Department of Energy will provide public access to these results of federally sponsored research in accordance with the DOE Public Access Plan  ( https://www.energy.gov/doe-public-access-plan ).}
\affiliation{Quantum Information Science Section, Oak Ridge National Laboratory, Oak Ridge TN 37831, USA}

\date{\today}

\begin{abstract}
Kitaev's honeycomb model is a paradigmatic exactly solvable system hosting a quantum spin liquid with non-Abelian anyons and topologically protected edge modes, offering a platform for fault-tolerant quantum computation.
However, real candidate Kitaev materials invariably include complex secondary interactions that obscure the realization of spin-liquid behavior and demand novel quantum computational approaches for efficient simulation. 
Here we report quantum simulations of a 22-site Kitaev honeycomb lattice on a trapped-ion quantum processor, without and with non-integrable Heisenberg interactions that are present in real materials. 
We develop efficient quantum circuits for ground-state preparation, achieving high accuracy with energy errors equivalent to an effective temperature $\approx 0.2$ (in units of the Kitaev interactions), consistent with the experimentally relevant spin-liquid regime.
Starting from these states, we apply controlled perturbations and measure time-dependent spin correlations along the system's edge. In the non-Abelian phase, we observe chiral edge dynamics consistent with a non-zero Chern number, a hallmark of chiral topological order, which vanishes upon transition to the Abelian toric code phase.
Extending to the non-integrable Kitaev-Heisenberg model, we find that weak Heisenberg interactions preserve chiral edge dynamics, while stronger couplings suppress them, signaling the breakdown of topological protection.
These simulations establish edge-mode dynamical simulations on quantum computers as a staple technique to understand the topological and chiral phases of matter in Hamiltonians beyond integrability.

\end{abstract}
\maketitle

\section{Introduction}

Topologically ordered phases of matter defy the conventional framework of spontaneous symmetry breaking and local order parameters, potentially enabling the construction of inherently fault-tolerant quantum computers, but also posing significant experimental, theoretical, and computational challenges.
The Kitaev honeycomb model~\cite{kitaev2006anyons} provides an exactly solvable example of a gapless quantum spin liquid (QSL) that serves as a parent phase for gapped topological orders. This model is approximately realized in spin-orbit-coupled Mott insulators \cite{jackeli}, and its variants offer a promising route to realizing non-Abelian anyons (emerging under broken time-reversal symmetry) for topologically protected QSL-based qubits \cite{klocke2024spinliquidbasedtopologicalqubits}. Candidate materials, including $ \alpha\textendash{}\text{RuCl}_{3}$, $\text{BaCo$_{2}$(AsO}_{4})_{2}$, as well as various iridates, are fiercely studied, some even exhibiting signatures of exotic QSL states~\cite{trebst2017, hermanns2018, kim2015, sears2015, banerjee2017, banerjee2018, banerjee2016, takagi2019, matsuda2025, kasahara2018, kasahara2018majorana, yokoi2021, bruin2022, singh2010, liu2011, choi2012, ye2012, comin2012, hwan2015, singh2012relevance, kitagawa2018spin, bahrami2019}, and are often described with a Kitaev Hamiltonian plus non-integrable interactions~\cite{chaloupka2010, jiang2011possible, chaloupka2013, rau2014, valenti_challenges, hermanns2018}.

\begin{figure*}[!htb]
    \centering
    \includegraphics[width=1\linewidth]{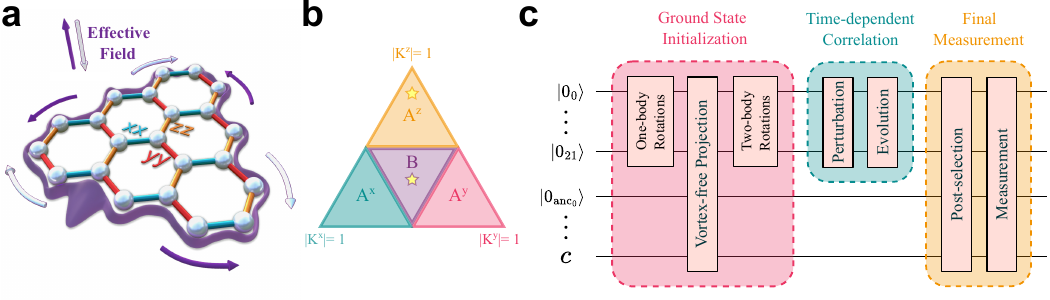}
    \caption{Model and quantum circuit overview. \textbf{a,} We simulate a 22-qubit Kitaev honeycomb lattice, featuring direction-dependent Ising interactions [see Eq.~(\ref{eq: kitaev})], an effective field with emergent three-body interactions [see Eq.~(\ref{eq: V})], and Heisenberg interactions [not shown; see Eq.~(\ref{eq: heis})]. Flipping a spin in the non-Abelian phase generates an edge excitation (purple bump) which travels clockwise or counterclockwise depending on the sign of the field. 
    \textbf{b,} Phase diagram of the Kitaev honeycomb in the absence of a magnetic field~\cite{kitaev2006anyons}. Stars indicate points we probe in the set of gapped Abelian phases $A^\alpha$ ($\alpha = x, y, z$) and the phase $B$, which acquires a gap and becomes a non-Abelian spin-liquid phase in the presence of a field.
    \textbf{c,} Quantum circuit schematic for measuring time-dependent spin-spin correlations along the edge.
    }
    \label{fig:Fig1}
\end{figure*}

Classical simulations have substantially advanced our understanding of Kitaev's model and its extensions, but they are fundamentally constrained especially in two or more dimensions: restricted system sizes tractable by exact diagonalization~\cite{rau2014, winter}, confinement to narrow cylindrical geometries in matrix product state methods~\cite{gohlke2017dynamics, barratt2021parallel, xiao2025}, restricted bond dimensions in (infinite) projected entangled-pair states~\cite{osorio2014probing}, and the sign problem encountered by quantum Monte Carlo~\cite{sato2021, nasu}. Quantum computers are a new and complementary technology for processing information stored in coherent quantum states, promising fundamentally better scaling for solving certain problems \cite{nielsen2010quantum} as well as new challenges for computer and algorithm development. They have recently been used to simulate topologically ordered systems \cite{edmunds2024constructingspin1haldanephase,kumaran2024transmon} including Kitaev model ground states \cite{xiao2021determining}, quasiparticle manipulation \cite{park2025digital}, a Kitaev-model substrate for emergent fermion simulations \cite{evered2025probing}, and a non-equilibrium Floquet Kitaev model with topological order \cite{will2025probing}. An important frontier is the application of quantum computers to realistic simulations of Kitaev-like materials, to probe experimental signatures of fractionalized quasiparticles intrinsic to topological quantum matter. Continuum features in inelastic neutron scattering \cite{banerjee2016,banerjee2017,banerjee2018} and proposed quasiparticle detectors \cite{klocke2021time, liu2022, halasz2024, klocke2024spinliquidbasedtopologicalqubits} require relatively complex quantum simulations, while the chiral edge transport leading to the quantized thermal Hall effect \cite{kasahara2018majorana,yokoi2021,bruin2022} is a natural first target.

We present and execute quantum circuits probing local chiral edge transport as a signature of global non-Abelian topological order in Kitaev-Heisenberg models, in both integrable and non-integrable limits.  Our results are based on three main advances.  First, we find that chiral edge correlations and other aspects of topological spin liquid behavior, as well as their breakdown at finite Heisenberg exchange, can be reliably captured in surprisingly small lattices of 22 spins. This small size is amenable to quantum computation, and the signatures of topological order here may also be of interest to the broader physics community. Second, we develop efficient algorithms for ground-state preparation and time evolution, using a hybrid protocol of measurement, feed-forward, and variational unitary ans\"atze \cite{bespalova2021quantum,umeano2025quantum}, enabled by recent advances in quantum hardware and algorithms \cite{iqbal2024topological,iqbal2024non}. These circuits accurately simulate the Abelian and non-Abelian phases of the Kitaev model, as well as a non-integrable regime with Heisenberg interactions that are important for modeling real candidate materials. Third, we execute these circuits on a trapped-ion quantum computer, providing a rigorous benchmark of current hardware capabilities. The results show excellent agreement with exact verification, demonstrating a new approach to understanding chiral and topological phases of strongly correlated materials through edge-mode dynamics in quantum simulations. 

\section{Model Hamiltonian} The exactly solvable Kitaev model on the honeycomb lattice~\cite{kitaev2006anyons} stabilizes a quantum spin liquid characterized by fractionalized quasiparticles and an emergent $\mathbb{Z}_{2}$ gauge structure. The model consists of bond-dependent Ising-like interactions between spin-$1/2$ degrees of freedom on 
a hexagonal lattice, as shown in Fig.~\ref{fig:Fig1}\textbf{a}. The bond Hamiltonian is
\begin{equation} \label{eq: kitaev}
    H_K = \sum_{\text{$x$\ bonds}}K^x \sigma_{j}^{x} \sigma_{k}^{x} + \sum_{\text{$y$\ bonds}}K^y \sigma_{j}^{y} \sigma_{k}^{y}  + \sum_{\text{$z$\ bonds}}K^z \sigma_{j}^{z} \sigma_{k}^{z}, 
\end{equation} 
where $K^\alpha$ with $\alpha \in \{x,y,z\}$ are bond-dependent coupling strengths, while $\sigma_{j}^{\alpha}$ are Pauli operators representing spin-$1/2$ degrees of freedom at the lattice sites $j$. Interactions on the three bonds around any given site do not commute, resulting in exchange frustration and a quantum spin liquid ground state. However, the Hamiltonian does commute with the plaquette, or vortex, operators:

\begin{equation}\label{eq: plaquette}
\def\hexscale{0.5}
W_p = \sigma_1^x \sigma_2^y\sigma_3^z\sigma_4^x\sigma_5^y\sigma_6^z, ~~~~~
\begin{tikzpicture}[scale=\hexscale,baseline={([yshift=-.5ex]current bounding box.center)}]
    \def\fontscale{\hexscale}
    \def\offset{0.4}
    
    \coordinate (0) at (0,0);
    \coordinate (3) at (0, 1);                      
    \coordinate (4) at ({sqrt(3)/2}, {0.5});  
    \coordinate (5) at ({sqrt(3)/2}, -{0.5});  
    \coordinate (6) at (0, -1);                     
    \coordinate (1) at (-{sqrt(3)/2}, -{0.5}); 
    \coordinate (2) at (-{sqrt(3)/2}, {0.5});  
    \coordinate (3z) at ({0},{2-\offset});                      
    \coordinate (4x) at ({sqrt(3)-\offset*sin(60))}, {1-\offset*cos(60)});  
    \coordinate (5y) at ({sqrt(3)-\offset*sin(60)}, -{1+\offset*cos(60)});  
    \coordinate (6z) at (0,-{2+\offset});                     
    \coordinate (1x) at (-{sqrt(3)+\offset*sin(60)}, -{1+\offset*cos(60)}); 
    \coordinate (2y) at (-{sqrt(3)+\offset*sin(60)}, {1-\offset*cos(60)});  

    \draw[thick] (3) -- (3z);        
    \draw[thick] (4) -- (4x);      
    \draw[thick] (5) -- (5y);     
    \draw[thick] (6) -- (6z);      
    \draw[thick] (1) -- (1x);      
    \draw[thick] (2) -- (2y);      
    \draw[thick] (3) -- (4) -- (5) -- (6) -- (1) -- (2) -- cycle;

    \node[circle, draw, fill=white, minimum size=1mm, scale=0.75*\hexscale] at (3) {};
    \node[circle, draw, fill=white, minimum size=1, scale=0.75*\hexscale] at (4) {};
    \node[circle, draw, fill=white, minimum size=1, scale=0.75*\hexscale] at (5) {};
    \node[circle, draw, fill=white, minimum size=1, scale=0.75*\hexscale] at (6) {};
    \node[circle, draw, fill=white, minimum size=1, scale=0.75*\hexscale] at (1) {};
    \node[circle, draw, fill=white, minimum size=1, scale=0.75*\hexscale] at (2) {};

    \node at (0, 0.9) [below,scale=\fontscale*1.25] {$3$};
    \node at (4) [below left,scale=\fontscale*1.25] {$4$};
    \node at (5) [above left,scale=\fontscale*1.25] {$5$};
    \node at (0,-0.9) [above,scale=\fontscale*1.25] {6};
    \node at (1) [above right,scale=\fontscale*1.25] {$1$};
    \node at (2) [below right,scale=\fontscale*1.25] {$2$};

    \node at (3z) [below right,scale=\fontscale*1.5] {$z$};
    \node at (4x) [above left,scale=\fontscale*1.5] {$x$};
    \node at (5y) [below left,scale=\fontscale*1.5] {$y$};
    \node at (6z) [above right,scale=\fontscale*1.5] {$z$};
    \node at (1x) [below right,scale=\fontscale*1.5] {$x$};
    \node at (2y) [above right,scale=\fontscale*1.5] {$y$};

\end{tikzpicture} 
    .
\end{equation}
The $\mathbb{Z}_{2}$ gauge fluxes correspond to these plaquette operators that have eigenvalues $\pm 1$. 

The general Kitaev Hamiltonian $H_K$ in Eq.~(\ref{eq: kitaev}) supports two distinct types of phases at zero temperature~\cite{kitaev2006anyons}, as shown in Fig.~\ref{fig:Fig1}\textbf{b}. When one coupling dominates, with $|K^\alpha| > |K^{\beta}| + |K^\gamma|$ for distinct $\alpha, \beta, \gamma$, the system enters an Abelian topological phase $A^\alpha$ equivalent to the toric-code model. In contrast, the gapless $B$ phase emerges when $|K^\alpha| \leq |K^{\beta}| + |K^{\gamma}|$ holds for all directions $\alpha, \beta, \gamma$, yielding a $\mathbb{Z}_2$ Dirac spin liquid. Introducing a time-reversal-symmetry-breaking perturbation, such as a weak magnetic field, gaps the $B$ phase and stabilizes a non-Abelian quantum spin liquid. This topologically ordered phase hosts Ising anyons---exotic quasiparticles that underpin Clifford operations in proposed fault-tolerant topological quantum computers~\cite{kitaev2003fault, bonderson2010}.

In Kitaev's original analysis \cite{kitaev2006anyons}, a low-energy effective Hamiltonian was derived for the system in a weak magnetic field by retaining the leading-order terms that preserve the plaquette flux sector $W_{p} = 1$. This perturbative expansion yields three-spin interactions in the bulk that open an energy gap and stabilize the non-Abelian topological phase. If the system also has a boundary, then certain boundary components of the magnetic field can be directly included in the Hamiltonian, preserving the exact solution and lifting unphysical degeneracies associated with ``dangling'' Majorana fermions~\cite{kao2024vacancy,kao2024dynamics,zhang2025low,zhang2025probing}. The resulting effective field Hamiltonian is
\begin{align} \label{eq: V} H_V = V \sum_{\substack{\text{adjacent}\\ \text{$x$, $y$\ bonds}}} \sigma^x_i \sigma^z_j \sigma^y_k + V \sum_{\substack{\text{adjacent}\\ \text{$x,z$\ bonds}}} \sigma^x_i \sigma^y_j \sigma^z_k \nonumber\\ + V \sum_{\substack{\text{adjacent}\\ \text{$y,z$\ bonds}}} \sigma^y_i \sigma^x_j \sigma^z_k  + h \sum_{\text{boundary}} \sigma_{j}^{\alpha}.\end{align} 
Here $h$ is the magnetic field strength on the boundary, the interaction direction $\alpha$ is given by the missing bond around a given boundary site $j$, while the emergent three-body operators are proportional to the commutators between adjacent bonds and have coupling strengths $V \sim h^3$. For simplicity, we treat $h$ and $V$ as independent tunable parameters, decoupling the boundary field strength from the induced three-body interactions.  This choice allows us to employ larger $V$, providing a crucial enhancement of the excitation gap and suppression of the correlation length, to clearly resolve the topological signatures on a quantum computer.

Candidate Kitaev materials are believed to display dominant Kitaev interactions as well as weaker symmetry-allowed interactions~\cite{symmetries}. These additional interactions break the integrability of Kitaev's model and pose significant challenges for conventional numerical methods. One important example is the Heisenberg interaction~\cite{rau2014} which we consider in the limit of small coupling $J$ \cite{valenti_challenges, eichstaedt2019deriving}
\begin{equation}\label{eq: heis}
    H_J = J 
    \sum_{\text{all\ bonds}} \vec\sigma_{j}\cdot \vec\sigma_{k}.
\end{equation}
Including all types of interactions in Eqs.~(\ref{eq: kitaev}), (\ref{eq: V}), and (\ref{eq: heis}) we have the total Hamiltonian 
\begin{equation} H = H_K + H_V + H_J. \end{equation}
The Hamiltonian $H_K + H_V$ is exactly solvable, while $H_J$ breaks the integrability.

A key feature of the Kitaev model with magnetic field is the existence of topologically protected chiral fermion edge modes \cite{kitaev2006anyons}. From the perspective of the emergent fermions in the Kitaev model, the two-body terms in $H_K$ correspond to first-neighbor hopping and generate Dirac point nodes as in graphene~\cite{castroneto2009}, while the three-body terms in $H_V$ are equivalent to second-neighbor hopping and gap these Dirac nodes into topological bands with finite Chern numbers~\cite{haldane1988}. By the bulk-edge correspondence, these topological fermion bands in the bulk then directly translate into chiral fermion edge modes that propagate in only one direction around the boundary~\cite{hasan2010}. Upon adding a Heisenberg interaction $J$, these chiral edge modes are expected to persist until the phase transition out of the QSL phase. While this phase transition has been studied for the Kitaev model without a magnetic field~\cite{zhang2021variational,chaloupka2013,gotfryd2017phase,gohlke2017dynamics}, it is not {\it a priori} clear how the magnetic field shifts the critical value of $J$. 

In the following sections, we develop an approach to probe chiral edge dynamics in instances of the Kitaev and Kitaev-Heisenberg models, using quantum circuits presented schematically in Fig.~\ref{fig:Fig1}\textbf{c}. The circuits combine state preparation that is optimized on conventional computers with scalable quantum time-evolution. With twenty-two spins we present a step in a viable route towards beyond-classical simulations on future quantum devices.

\section{Initial state circuits} Our first step is to prepare ground states of $H$ on a quantum computer. The most well-known circuits for ground state preparation use local variationally-optimized unitary ans{\"a}tze, however, these require extensive depths to create long-range topological entanglement in quantum spin liquids \cite{tantivasadakarn2023hierarchy}.  Recent efficient alternatives have used deterministic non-unitary dynamics, based on mid-circuit measurements and measurement-outcome-dependent ``feed-forward" operations,  with promising results obtained for the toric code \cite{iqbal2024topological}, states with $D_4$ topological order \cite{iqbal2024non}, and low-energy states of the  Kitaev model \cite{evered2025probing}. Here we adapt these protocols to determine high-fidelity circuits for the present model. 

Our approach for preparing Kitaev ground states, with Heisenberg coupling $J=0$, is summarized in Fig.~\ref{fig:Fig1}\textbf{c}; we use a related approach for nonzero $J$ at the end of this section. Beginning from  an initial state $\ket{\bm0}=\bigotimes_i \ket{0_i}$, we prepare an approximate Kitaev ground state 
\begin{equation} \label{psi0} \ket{\psi_0(\bm \varphi,\bm\theta)} = U_2(\bm \varphi) P_\text{VF} U_1(\bm \theta)\ket{\bm 0} \end{equation}
using parameterized single-qubit rotations $U_1(\bm \theta)$, the projection to the vortex-free sector $P_\text{VF} \propto \prod_p (1+W_p)$ enabled by measurement and feed-forward, and problem-inspired many-body rotations $U_2(\bm \varphi)$. We now summarize each of these components (see Appendix \ref{initial states appendix} for more technical details). 

The first unitary $U_1(\bm \theta)$ performs single-qubit rotations that serve two purposes.  First, for future use in feed-forward, it initializes six ``correction qubits" $c_p$  on the boundary (Supplemental Data Fig.~\ref{fig:GS_Opt}) in eigenstates of Pauli operators $P_{c_p}$ that anticommute with the corresponding plaquette operators $W_p$.  Second, it rotates all other qubits to maximize the ground-state fidelity after projection to the vortex-free sector.

Next, we use measurement and feed-forward to apply the non-unitary projector $P_\text{VF}$. Similar to Ref.~\onlinecite{iqbal2024topological}, we entangle each plaquette $p$ with an ancilla $a$, such that measuring the ancilla collapses the state into the $W_p=1$ or $W_p=-1$ eigensector, depending on the ancilla measurement outcome. The projection circuit is structured as

\begin{equation}\label{eq: plaquette-circuit}
\vcenter{\hbox{\scalebox{0.75}{%
\begin{quantikz}[row sep={0.55cm,between origins}, column sep=0.32cm]
  \lstick{$\ket{0}_{a_p}$} & \gate{H} & \ctrl{1}\gategroup[7,steps=6,style={dashed,rounded
      corners,inner xsep=2pt,inner ysep=4pt},background,label style={label
      position=below,anchor=north,yshift=-0.25cm}]{{Controlled-$W_p$}}
    & \ctrl{2} & \ctrl{3} & \ctrl{4} & \ctrl{5} & \ctrl{6} & \gate{H} & \meter{} \\
  \lstick{$\sigma_1$} & & \gate{X}    & & & & & & & \\
  \lstick{$\sigma_2$} & & & \gate{Y} & & & & & & \\
  \lstick{$\sigma_3$} & & & & \gate{Z} & & & & & \\
  \lstick{$\sigma_4$} & & & & & \gate{X} & & & & \\
  \lstick{$\sigma_5$} & & & & & & \gate{Y} & & & \\
  \lstick{$\sigma_6$} & & & & & & & \gate{Z} & &
\end{quantikz}}}}
\end{equation}
following Eq.~(\ref{eq: plaquette}). Plaquettes with $W_p=1$ are in the correct vortex-free sector, while plaquettes with $W_p=-1$ are corrected by applying the correction Pauli operator $P_{c_p}$; see Appendix \ref{initial states appendix} for additional details. Crucially, this approach always produces the same state, irrespective of the measurement results, providing a universal starting point for our next step.

Finally, we use many-body rotations $U_2(\bm \varphi)$ to evolve to the ground state.  We choose $U_2(\bm \varphi)$ based on a $d$-layer Hamiltonian variational ansatz~\cite{wecker2015progress}. In the Kitaev limit ($J=0$), $U_2(\bm \varphi)$ is a product of exponentials of the one- and two-body terms in $H_K$ and $H_V$, which keeps the state within the vortex-free sector. 

We find the ansatz in Eq.~(\ref{psi0}) to be highly effective, with numerical parameter optimizations of the fidelity (computed using a conventional computer)
\begin{equation}\label{eq:F}
    F = \big|\langle \psi_\text{GS} \ket{\psi_0(\bm \varphi,\bm\theta)}  \big|^2,
\end{equation}
yielding rapid convergence to the ground state $\ket{\psi_\text{GS}}$ as layers $d$ are added to $U_2(\bm \varphi)$ (Supplemental Data Fig.~\ref{fig:GS_Opt}). The circuits obtain numerical fidelities greater than 0.98 in the Kitaev limit $J=0$, using 108 two-qubit gates in $U_2(\bm \varphi)$ in the Abelian phase and 135 two-qubit gates in the non-Abelian phase. We validate these on a quantum computer in the next section.

Heisenberg couplings $|J|>0$ introduce new challenges, as $\comm{W_p}{H_J}\neq 0$, and therefore the ground states are superposed over multiple plaquette eigensectors. We consider state preparation with $J=0.05$ and $J=0.2$ in our results. For the weaker coupling we expect the state to be close to the pure Kitaev ground state, and find that our previous procedure, augmented with single-qubit rotations in $U_2(\bm \varphi)$ to drive transitions outside the vortex-free sector, succeeds in producing high-fidelity ground states.  With the stronger coupling there is significant amplitude outside the vortex-free sector, so we skip the vortex-free projection entirely. We instead begin in an initial state with single-qubit Pauli $\sigma^y$ rotations on each qubit, then apply $U_2(\bm \varphi)$ including single-qubit gates to drive transitions across vortex sectors, similar to the previous case.  For each choice of $J$ our final circuits contain 81 two-qubit gates in $U_2$, with numerical fidelities $F > 0.9$ indicating a close approximation to the ground state even in these nonintegrable cases.

\section{Quantum computer state preparation}

We now consider circuits executed on the quantum computer.  To mitigate computational errors in the Kitaev limit ($J=0$), we implement an error detection scheme based on measuring plaquette operators at the end of each circuit using Eq.~(\ref{eq: plaquette-circuit}) (Fig.~\ref{fig:Fig1}\textbf{c}). A plaquette violation indicates the presence of an error, and we examine the improvement that is gained by discarding results with one or more plaquette violations; see Ref.~\onlinecite{nigmatullin2024} for an alternative approach.

\begin{figure*}[!tb]
    \centering
    \includegraphics[width=\linewidth]{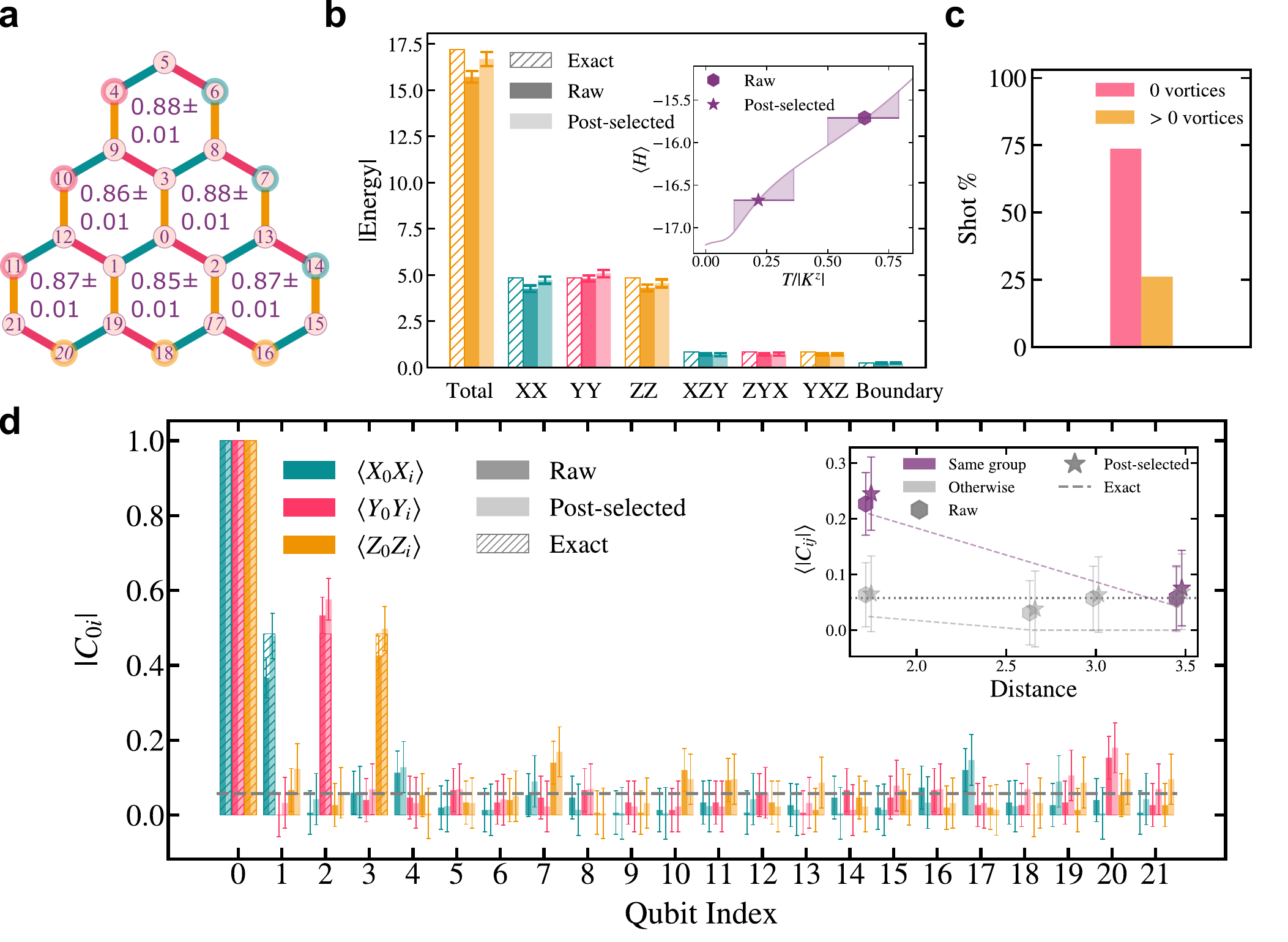}
    \caption{ Quantum state preparation. 
    \textbf{a,} Non-Abelian ground state preparation on the H2-1 quantum computer yields plaquette expectations close to the ideal vortex-free limit $\langle W_p\rangle = 1$. The Hamiltonian parameters are $K^x=K^y=K^z=-1$, $V=0.3$, $h=0.1$, and $J=0$.
    \textbf{b,} Term-resolved energy measurements ($\pm$ standard error) for the non-Abelian ground state. Post-selecting significantly improves agreement with the exact ground state energy. (inset) Effective temperature assignment by matching the measured state energy to the thermal $\langle H\rangle(T)$ curve (see Appendix \ref{temperature appendix}).
    \textbf{c,} Percentage of shots with and without vortices, for 300 shots.
    \textbf{d,} $\langle XX \rangle$, $\langle YY \rangle$, and $\langle ZZ \rangle$ correlations with the center-most qubit 0. Magnitudes are vanishing with respect to the shot-noise threshold (gray dashed line), with distinguishable exceptions for $\langle Z_{0}Z_{3} \rangle$, $\langle X_{0}X_{1}\rangle$, and $\langle Y_{0}Y_{2} \rangle$, confirming the ultra-short-range correlation property expected for the Kitaev model. (inset) Correlations among edge sites with only two bonds, associated with ``dangling" Majorana fermions in Kitaev's exact solution, show longer range correlations. Purple data is averaged over sites  within individual groups of identically colored edge sites in Fig.~2\textbf{a} ($XX$, $YY$, and $ZZ$ correlations within the teal, pink, and orange group of sites, respectively), while gray data is averaged over all other edge correlations. Unlike bulk qubits, edge qubits have correlations beyond nearest neighbors. Horizontal dotted gray line indicates shot-noise threshold.
    }
    \label{fig:energy}
\end{figure*}

We test our approach to initial state preparation and error mitigation on the Quantinuum H2-1 trapped-ion quantum computer. The plaquette operator expectation values are shown in Fig.~\ref{fig:energy}\textbf{a} for a state in the non-Abelian phase. In Fig.~\ref{fig:energy}\textbf{b}, we examine the accuracy of the expected energy for each term in the Hamiltonian. We observe reasonable agreement between the raw results and the exact expectation for each Hamiltonian term. In the raw results, 74\% of them contained zero plaquette violations, as shown in Fig.~\ref{fig:energy}\textbf{c} (see Supplemental Data Fig.~\ref{fig:vortices} for time-evolved behavior). Post-selecting to remove plaquette violations significantly improves the energy agreement while maintaining a large statistical sample. Similar results are shown for the Abelian phase in Supplemental Data Fig.~\ref{fig:abelian-energy}. 

\begin{figure*}[!htb]
    \centering
    \includegraphics[width=1\linewidth]{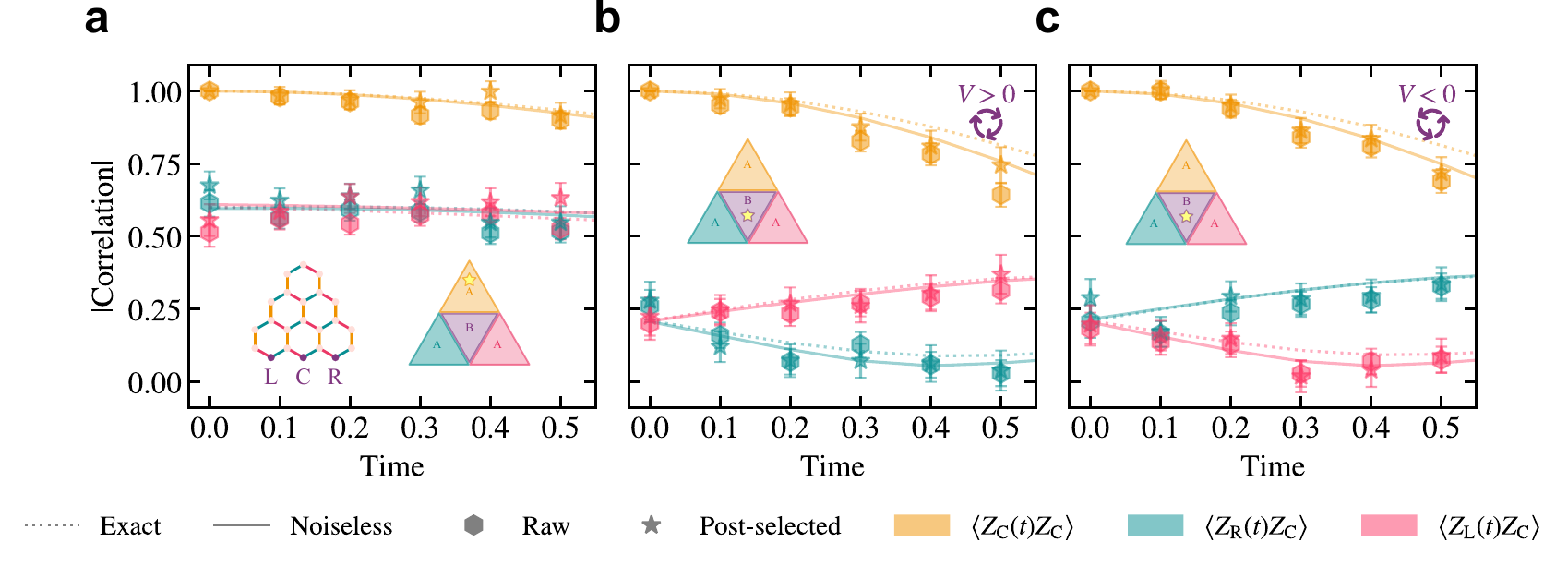}
    \caption{ Chiral edge dynamics in the pure Kitaev model $(J = 0)$.
    \textbf{a,} Time evolution of unequal-time spin-spin correlation functions $\langle Z_i(t) Z_\mathrm{C}\rangle$ on the edge in the Abelian phase, following a local excitation at reference site $\mathrm{C}$ (see inset for qubit labels). We observe no chiral response, consistent with a vanishing Chern number and the absence of topologically protected edge states. Parameters are $K^x=K^y=K^z/6=-1/6$, $V=0.3$, $h=0.1$, and $J=0$. 
    \textbf{b, c, } We apply the same protocol to a non-Abelian state with $V>0$ in \textbf{b}, and $V<0$ in \textbf{c}. The directional propagation observed in $\langle Z_{\mathrm{L}}(t) Z_{\mathrm{C}} \rangle$ and $\langle Z_{\mathrm{R}}(t) Z_{\mathrm{C}} \rangle$ reflects the presence of a chiral edge mode and is a dynamical signature of the nonzero Chern number and non-Abelian topological order. Results obtained on quantum hardware, both raw and post-selected, show quantitative agreement with an exact theoretical simulation (dotted line) and a noiseless circuit simulation that includes all algorithmic errors (solid line). We observe a (counter-)clockwise chiral behavior for the ($V<0$) $V>0$ case. Parameters are the same as the Abelian case except that $K^x=K^y=K^z=-1$.} 
    \label{fig:fig3}
\end{figure*}

It is interesting to consider what temperature would produce the energy we observed. Our quantum states are not thermal states, but understanding the magnitude of their energy errors in terms of an ``effective temperature" provides a loose contextualization of the errors in connection to condensed matter physics. In Appendix \ref{temperature appendix} we show how to match the measured energy with the thermal-average energy expected from the canonical ensemble. From this we obtain an effective temperature $T_{\text{eff}}\approx 0.2 |K^z|$ for the prepared state, as shown in the inset of Fig.~\ref{fig:energy}(b). For thermal states at this temperature, fractionalized excitations are already expected to dominate~\cite{nasu2015thermal}. Moreover, for the typical values of the Kitaev interaction in $\alpha$-RuCl$_3$---which are $1.5$ meV $\leq K \leq$ $5$ meV in our convention of representing the spins~\cite{valenti_challenges,eichstaedt2019deriving, banerjee2016,banerjee2017,yadav2016kitaev}---this temperature translates to a range between $3.8$ K and $12.5$ K. This temperature range overlaps with the physical temperatures in current thermal-transport experiments that also aim to probe chiral edge transport~\cite{kasahara2018majorana,yokoi2021,bruin2022}. 

Finally, we examine the bulk and edge correlations in Fig.~\ref{fig:energy}\textbf{d}. Inspecting the center-most bulk qubit (Fig.~\ref{fig:energy}\textbf{a}) against all other sites reveals that only $\langle Z_{0}Z_{3} \rangle$, $\langle X_{0}X_{1}\rangle$, and $\langle Y_{0}Y_{2} \rangle$ distinguishably survive above the shot-noise threshold (dashed gray line). This confirms the ultra-short-range correlation property characteristic of the Kitaev model~\cite{baskaran2007}.
For the gapped non-Abelian phase, edge correlations are theoretically expected to develop an algebraic decay in distance~\cite{kitaev2006anyons, willans2010}. While the exact ground state does not display this algebraic profile at the current system size, the inset of Fig.~\ref{fig:energy}\textbf{d} shows a sizable correlation for the first point beyond nearest neighbors within the highlighted edge groups (Fig.~\ref{fig:energy}\textbf{a}). This edge correlation is accurately captured in the quantum computer results, marking a clear departure from the bulk qubits where such correlations remain suppressed, modulo qubits 7, 17, and 20 (italic in Fig.~\ref{fig:energy}\textbf{a}) which are exhibiting anomalous correlations that are slightly exceeding the shot-noise floor.

Overall the modest plaquette violation probabilities, post-selection to the vortex-free sector, measured energy, low effective temperature, and correlation structure all strongly support that our circuits succeed in preparing approximate quantum spin liquid ground states on the quantum computer. 
 
\section{Time evolution circuits} To observe chiral edge dynamics, we will need to evolve our quantum states in time.  Typically this is performed using low-order Trotter decompositions of the propagator $\exp(-i H \tau)$, however, the large number of three-body interactions in our model make this impractical.  Instead we used the problem structure to derive an efficient approximation to the propagator, with controllable error. The same basic idea has been derived independently in Ref.~\onlinecite{kalinowski2023non}.

We begin by considering two-qubit Kitaev interactions along a single type of bond $\alpha \in \{x,y,z\}$, described by the unitary $U_\alpha(\tau) = \exp(-i \tau K^\alpha \sum_{\alpha\ \text{bonds}} \sigma^\alpha_j \sigma^\alpha_k)$. Sequentially implementing bond unitaries leads to a first-order Trotter approximation $U_x(\tau)U_y(\tau)U_z(\tau)$ of the evolution under $H_K$, with higher-order errors described by the Baker-Campbell-Hausdorff (BCH) formula $e^Ae^B = e^{A+B+[A,B]/2+\ldots}$ \cite{hall2013lie}. A key insight is that the leading-order error terms in the Trotter approximation are proportional to the three-body terms in $H_V$. Therefore, evolving under the two-body terms in $H_K$ is sufficient to implement the three-body interactions. 

A limitation of the ansatz $U_x(\tau)U_y(\tau)U_z(\tau)$ is that it does not provide independent control of the three-body interaction coefficients, and results in a mixture of coefficient signs.  To control the strength and signs of the three-body interactions we combine two sequences of bond interactions as
\begin{equation}
\label{time_evol_ansatz}
U(\tau) = \prod_{l=1}^2 U_\alpha(t_\alpha^{(l)})U_\beta(t_\beta^{(l)})U_\gamma(t_\gamma^{(l)}), 
\end{equation}
where $\alpha \neq \beta \neq \gamma \in \{x,y,z\}$.  In Appendix \ref{time evolution appendix}, we use the BCH formula to derive parameters for which $U(\tau) = \exp(-i (H_K + H_V) \tau + \ldots)$ with a chosen $V$ and $h=0$, where  $\ldots$ denotes errors at third order in the exponent (see Supplemental Data Fig.~\ref{various_dts} for convergence). A non-zero boundary field $|h|>0$ and Heisenberg coupling $|J|>0$ are included through standard Trotterization, yielding a controlled approximation to $\exp(-i H \tau)$.

While the analytic theory of the propagator $U(\tau)$ provides a useful starting point for the ansatz parameters, we have found it is possible to increase the accuracy using an efficient numerical optimization. For this we consider a unitary $U'(\tau)$ and Hamiltonian $H'$ on a T-junction, which is defined to contain four adjacent spins with bonds $\sigma^x_i\sigma^x_j$, $\sigma^y_i\sigma^y_k$, and $\sigma^z_i\sigma^z_l$ along with the three-body interactions acting only on those spins.  Minimizing the 2-norm error $||e^{-i H' \tau} - U'(\tau)||_2$ determines optimized parameters for $U'(\tau)$.  These almost exactly account for higher-order errors in the BCH expansion, since there are only six BCH terms on the T-junction and we have six free parameters in (\ref{time_evol_ansatz}) to optimize them. We then apply $U(\tau)$ with the T-junction-optimized parameters on our whole lattice, accounting for some of the higher-order BCH errors on the lattice and modestly improving the accuracy as shown in Supplemental Data Fig.~\ref{fig:optimized parameters}.

\section{Chiral edge dynamics}

With all the circuit components ready, we now move to the major result of this work: detection of the chiral edge dynamics. Our approach is to compute a two-time correlation function $\langle Z_i(t)Z_\mathrm{C} \rangle$ between the purple edge qubits in Fig.~\ref{fig:fig3}\textbf{a}. This correlation function physically corresponds to acting with operator $Z_{\mathrm{C}}$ on the central qubit $\mathrm{C}$ to set up a localized excitation packet, then after time $t$ measuring local magnetizations $Z_{\mathrm{L}}$ and $Z_{\mathrm{R}}$ on its left and right to detect preferential excitation propagation in either direction along the edge. The chiral edge dynamics manifests in a marked asymmetry between the resulting time-dependent two-point correlations, $\langle Z_{\mathrm{L}}(t)Z_{\mathrm{C}}\rangle$ and $\langle Z_{\mathrm{R}}(t)Z_{\mathrm{C}}\rangle$, with one increasing and the other decreasing in time.

We compute the time-dependent correlations using the direct measurement approach of Ref.~\onlinecite{mitarai2019} due to its efficient scaling and previously demonstrated performance~\cite{eassa2024}. This requires executing three types of circuits: a real-part circuit with mid-circuit measurement of $M=Z_{\mathrm{C}}$, and two imaginary-part circuits with $R^z_{\mathrm{C}}(\pm\pi/4)$ gates acting on $\mathrm{C}$, all before the time evolution. 
Their results are combined to produce 
\begin{equation}\label{eq: direct}
\begin{aligned}
    \langle Z_i(t)Z_{\mathrm{C}}\rangle = p_+\langle Z_i\rangle_{M_+}
    - p_-\langle Z_i\rangle_{M_-}
    - \frac{i}{2}(\langle Z_i \rangle_+ - \langle Z_i \rangle_-),
\end{aligned}
\end{equation}
where $p_\pm$ are mid-circuit measurement probabilities of $Z_\text{C}=\pm 1$, while the final expectation values $\langle Z_i \rangle_\pm$ are computed from the imaginary-part circuits with corresponding signs in $R^z_{\mathrm{C}}(\pm\pi/4)$. Note that while the physical correlation is bounded as $|\langle Z_i(t)Z_{\mathrm{C}}\rangle | \leq 1$, the quantum computational estimates of Eq.~(\ref{eq: direct}) are based on measurements from three separate circuits,  and the shot noise in these measurements can produce mean values or error bars for which the correlation exceeds unity.

We investigate the evolution of the time-dependent correlations starting from Abelian (Fig.~\ref{fig:fig3}\textbf{a}) 
and non-Abelian (Fig.~\ref{fig:fig3}\textbf{b, c})
ground states, computed using 750 shots for each circuit.  
In the non-Abelian case, the correlation magnitude $|\langle Z_{\mathrm{L}}(t)Z_{\mathrm{C}}\rangle|
$ increases in time with a corresponding decrease in $|\langle Z_{\mathrm{R}}(t)Z_{\mathrm{C}}\rangle|$ in Fig.~\ref{fig:fig3}\textbf{b}, consistent with the presence of a clockwise chiral edge current for $V>0$. For $V<0$ (Fig.~\ref{fig:fig3}\textbf{c}), we observe the opposite behavior, suggesting a counter-clockwise chiral edge current instead. This is in contrast to the Abelian case, in which the cross-correlations show approximately uniform behavior.  The results are in very good agreement with the noiseless circuit simulations, including all sources of algorithmic error, as well as exact simulations of the physics. They are consistent with the presence of a topologically protected edge mode in the non-Abelian phase, as well as the absence of such an edge mode in the Abelian phase, as expected \cite{kitaev2006anyons}.

Curiously, unlike the initial state energy measurements in Fig.~\ref{fig:energy}\textbf{a}, the post-selection does not appear to significantly change most of the results in Fig.~\ref{fig:fig3}. This may be due to 
the increased error probability at larger depths (Supplemental Data Fig.~\ref{fig:vortices}), resulting in a small statistical sample with non-negligible fluctuations, or to the robustness of the observed behavior to quantum computational errors.

 \begin{figure}[!htb]
    \includegraphics[width=0.9\columnwidth]{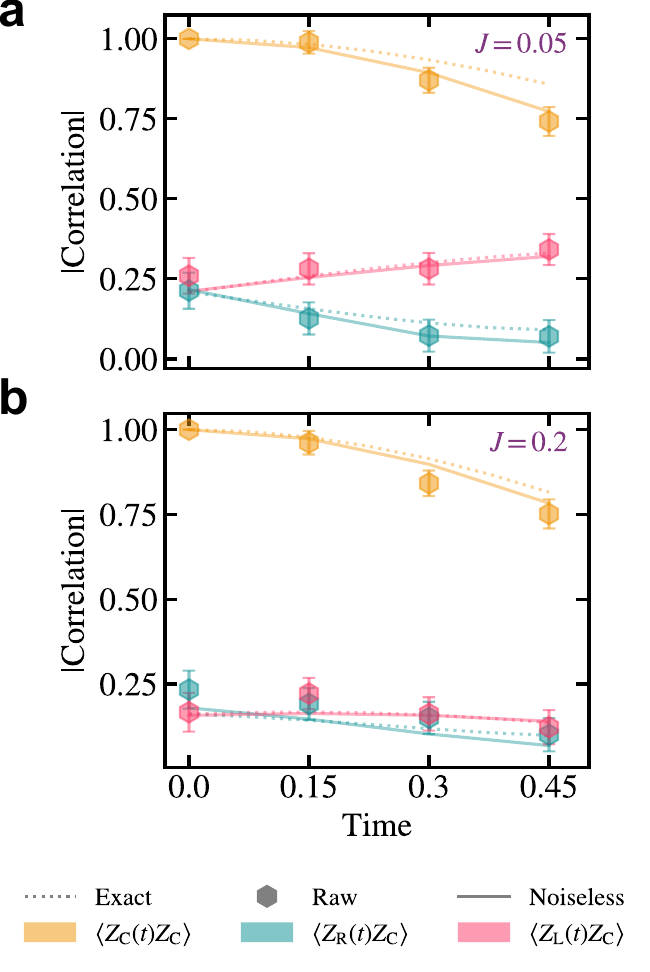}
    \caption{Chiral dynamics in the Kitaev-Heisenberg model. \textbf{a,} Starting from the ground state of the Kitaev-Heisenberg model with $J = 0.05$, a local spin excitation is applied to site $\mathrm{C}$ and unequal-time correlations are measured, similar to Fig.~\ref{fig:fig3}. The observed directional propagation is consistent with a nonzero Chern number and the persistence of non-Abelian topological order in the presence of weak Heisenberg interactions. \textbf{b,} For a stronger Heisenberg coupling $J = 0.2$, the chiral dynamics is strongly suppressed. Both panels use $K^{\alpha} = -1 \; (\alpha = x, y, z)$, $V=0.3$, and $h=0.1$.
    }
    \label{fig:Fig4}
\end{figure}

Next, we explore the non-integrable limit by adding the Heisenberg interaction $J$ in Eq.~(\ref{eq: heis}). We focus on two values for $J$, a weak perturbative $J = 0.05$, and a stronger $J = 0.2$. The energies of their prepared ground states are shown in Supplemental Data Fig.~\ref{fig:heis_energies}. The Kitaev-Heisenberg time evolution circuits require additional two-qubit gates that were not present in the pure Kitaev model to implement the Heisenberg interactions.  This doubles the two-qubit gate count for each timestep (see Supplemental Data Table~\ref{table: gates} for a recap of the two-qubit gate counts).  We therefore choose a larger timestep of $\tau=0.15$, to extend these simulations to a time that is comparable to the pure Kitaev case while maintaining a similar gate count. 

Fig.~\ref{fig:Fig4} shows the resulting behavior. Our circuits capture the exact behavior with great accuracy. The chiral edge current is stable with respect to weak $J$ perturbations at $J = 0.05$ (Fig.~\ref{fig:Fig4}\textbf{a}), however, increasing the strength of the Heisenberg interaction to $J= 0.2$ results in major suppression of the chiral behavior (Fig.~\ref{fig:Fig4}\textbf{b}). This transition in the chiral edge properties between $J = 0.05$ and $J = 0.2$ is consistent with the phase transition out of the Kitaev QSL phase previously found in the Kitaev-Heisenberg model~\cite{zhang2021variational,chaloupka2013,gotfryd2017phase,gohlke2017dynamics}. Note there is no true phase transition in a finite system and---even for an infinite system---the magnetic field may shift the critical $J$ value. Nevertheless, our results show that the chiral edge current is maintained in the presence of a weak Heisenberg coupling and suppressed for a sufficiently large coupling, consistent with the transition to a qualitatively different ground state in the presence of non-negligible Heisenberg interactions.  

\section{Conclusions}

We showcased a framework for probing topological order of candidate Kitaev materials by developing and implementing efficient quantum circuits to prepare approximate ground states of a 22-spin Kitaev honeycomb lattice in an effective field, without and with realistic Heisenberg interactions, and observing their time-dependent chiral edge correlations using efficient time-evolution circuits.
The ground state circuits combined measurement, feed-forward, and variational circuit optimization to prepare highly-accurate approximations to the ground state, as verified by measuring energy errors, correlations, and relating them to an effective temperature.
The effective temperature in the Kitaev non-Abelian phase was similar to temperatures employed in real experiments targeting thermal edge transport in materials such as $\alpha$-RuCl$_3$, and the temperature methodology is applicable to Kitaev simulations on larger lattices.
We then probed chiral edge dynamics of the model using two-time correlation functions along the edge.  
These verified chiral signatures in the Kitaev non-Abelian phase and their absence in the Kitaev Abelian phase, as expected from Kitaev's original analysis \cite{kitaev2006anyons}, as well as characterizing how the dynamics are affected by Heisenberg couplings of varying strengths as are present in real materials. 
The results were in very good agreement with classical verification, validating the methodology and readiness to scale to larger lattices and timescales, where they can begin to connect with real condensed matter experiments.

Scaling our approach will require ground state circuits for non-integrable extensions of the Kitaev model at much larger sizes. We determined these circuits using numerical routines that were optimized against exact ground state calculations, which could be implemented at about twice the current size on modern supercomputers.  Tensor-network-based simulations offer appealing alternatives for compressed state representations that can model much larger systems; they have already been shown to be effective in representing and studying ground states of Kitaev-Heisenberg models~\cite{jiang2011possible, osorio2014probing, gohlke2017dynamics} and their use in synthesizing quantum circuits for many-body state preparation is an active area of research~\cite{berezutskii2025tensor, rudolph2023decomposition, dborin2022matrix, rogerson2024quantum, anselme2024combining, lin2021real, robertson2023approximate, gibbs2025learning}. It may also be possible to seed larger circuits with parameters from optimizations on smaller instances, to reduce the computational expense \cite{yu2023simulating}.  Additional alternatives include initializations to nearby ground states derived from Kitaev's analytic solution \cite{kalinowski2023non,xiao2021determining,park2025digital}, adiabatic state preparation~\cite{farhi2000quantum}, Pauli path simulation~\cite{lin2026}, or variational circuit optimization based on energy measurements from a quantum device \cite{cerezo2021variational}.   Once a state preparation circuit has been identified, then scalable quantum dynamics can be simulated on a quantum computer, similar to the present work.  These dynamics can generate highly-entangled states that cannot be addressed by conventional computing methods alone.  Further research along this path may thereby achieve beyond-classical computation of non-equilibrium topological phenomena in realistic models of materials.

With future advances in quantum computation, we can imagine several promising extensions of the current approach towards modeling physics of candidate Kitaev materials.  One direction is quantitative simulations of thermal transport experiments that directly probe the quantized thermal Hall effect \cite{kasahara2018majorana,yokoi2021,bruin2022}. Their temperature dependence can be explored using quantum computers by drawing on ideas from contemporary quantum thermodynamics, to engineer quantum states with various energies and effective temperatures  \cite{deutsch2018eigenstate,popescu2006entanglement,lotshaw2019simulating,lotshaw2021asymmetric,lotshaw2018quantum}. Another possible direction is to use more sophisticated techniques to set up and detect a wave packet traveling along the edge, to realize the time-domain anyon interferometry introduced in Ref.~\onlinecite{klocke2021time} as a model for future quasiparticle detectors. Finally, extending the current simulations to calculate dynamical spin structure factors would provide a connection to inelastic neutron scattering experiments \cite{banerjee2016,banerjee2017,banerjee2018}, to explore distinct dynamical signatures of topological phases of matter.

\section*{Acknowledgements}
The authors thank Eugene Dumitrescu for feedback on the manuscript, as well as Drew Potter,  Henrik Dreyer, Matthew DeCross, and David Amaro for feedback and suggesting to quantify energy using an effective temperature. A.A. thanks Norhan M. Eassa for insightful discussions.  This work was primarily led and supported by the U.S. Department of Energy, Office of Science, National Quantum Information Science Research Centers, Quantum Science Center,  for project conception, theory development, classical and quantum simulations, analysis, and writing.  J.G. was supported with funding from AWE.  This research used resources of the Oak Ridge Leadership Computing Facility at the Oak Ridge National Laboratory, which is supported by the Office of Science of the U.S. Department of Energy under Contract No. DE-AC05-00OR22725.

\appendix

\section{Trapped Ion Experiments}
All experiments were performed using Quantinuum's H2-1 device, except for Fig.~\ref{fig:fig3}\textbf{c}, which was done using H2-2. 300 shots were used for each of the energy measurements, while 750 shots were used for the time-dependent correlations. The zeroth time point in the correlation figures was computed from the ZZ energy measurements; the other points were computed using the direct-measurement method, as explained in the main text. The zeroth point in Fig.~\ref{fig:fig3}\textbf{c} was redone due to the initial trial having a significant fluctuation in $\langle Z_{\mathrm{R}}(t)Z_{\mathrm{C}}\rangle$. The two-qubit gate counts for each experiment are summarized in Supplemental Data Table~\ref{table: gates}.

\section{Initial State Circuits} \label{initial states appendix}
\subsection{Ansatz}
Our ground state preparation (Eq.~(\ref{psi0})) begins with single-qubit rotations
\begin{equation}
U_1(\bm \theta) = \prod_{q=1}^N  R^z_q(\theta_{q}^{z}) R^y_q(\theta_{q}^{y}),\end{equation}
where $R^\alpha_q(\theta^\alpha_q) = \exp(i\theta \sigma^\alpha_q)$. The rotations fix the initial state of the correction qubits (dark purple qubits on the boundary in Fig.~\ref{fig:GS_Opt}\textbf{a} (inset)) to $\ket{0}$, apart from qubit $\mathrm{C}$ in the bottom center of the lattice, which begins in $\ket{+}$.

Projection to the vortex-free sector is accomplished as follows.  For each plaquette $p$ we apply a Hadamard gate to an ancilla qubit, followed by six ancilla-controlled-Pauli gates that combine to yield a controlled-$W_p$, then apply another Hadamard to the ancilla.  This yields a state 
\begin{equation} \ket{\psi} = \frac{\ket{0_a}(1+W_p)U_1(\bm \theta)\ket{\bm 0} +\ket{1_a}(1-W_p)U_1(\bm \theta)\ket{\bm0}}{2}\end{equation}
Measuring the ancilla produces a state $\sim (1\pm W_p)U_1(\bm \theta)\ket{\bm 0}$, where the sign in $(1\pm W_p)$ is random and depends on the ancilla measurement outcome.  To obtain the desired $P_\text{VF} \propto \prod_p(1+W_p)$, we correct any operators $(1-W_p)$ by applying $P_{c_p}$ to the corresponding correction qubit. This yields $P_{c_p}(1-W_p)U_1(\bm \theta)\ket{\bm 0} = (1+W_p)U_1(\bm \theta)\ket{\bm 0}$, since by design $P_{c_p}$ anticommutes with $W_p$, and $U_1(\bm \theta)\ket{\bm 0}$ is an eigenstate of $P_{c_p}$ for the chosen qubit initialization. Iterating over all plaquettes, we deterministically obtain the vortex-free state $P_\text{VF} U_1(\bm \theta) \ket{\bm 0}$. 

In the Kitaev limit $J=0$ we evolve unitarily from the vortex-free state to the ground state using many-body rotations based on the Kitaev bond interactions 
\begin{equation} u_{j,k,d}^\alpha(\bm \varphi) = \text{exp}(-i\varphi_{j,k,d} \sigma_{j}^{\alpha} \sigma_{k}^{\alpha})\end{equation} 
Combining these we obtain our Hamiltonian variational ansatz
\begin{align}\label{eq:ansatz_1}
U_2(\bm \varphi) = & \prod_d \bigg(\bigg[\prod_{\alpha \in \{x,y,z\}}\prod_{\alpha \ \text{bonds}} u_{j,k,d}^\alpha(\bm \varphi)\bigg] \nonumber\\
& \times \prod_{\text{boundary}} \text{exp}(-i\varphi_{j,d}\sigma_{j}^{\alpha})\bigg). 
\end{align} 

To perform the ground-state preparation with nonzero Heisenberg coupling $J$, we use a similar ansatz design, except that after every two qubit gate, we add an extra parameterized single qubit rotation on both qubits
\begin{equation} \tilde u_{j,k,d}^\alpha(\bm \Phi) = \prod_{i=j,k} R_i^z(\Phi_{\alpha,i,d}^{(z)})R_i^y(\Phi_{\alpha,i,d}^{(y)})\end{equation}
This results in the generalized ansatz
\begin{align}\label{eq:ansatz_2}
U_2(\bm \varphi,\bm \Phi) = & \prod_d \bigg(\bigg[\prod_{\alpha \in \{x,y,z\}}\prod_{\alpha \ \text{bonds}} \tilde u_{j,k,d}^\alpha(\bm \Phi) u_{j,k,d}^\alpha(\bm \varphi) \bigg]\nonumber\\
& \times \prod_{\text{boundary}} \text{exp}(-i\varphi_{j,d}\sigma_{j}^{\alpha})\bigg)
\end{align} 
in place of the previous $U_2(\bm \varphi)$. The generalized ansatz retains a problem-inspired design with the exponentials of one- and two-body Kitaev terms of the Hamiltonian, with new hardware-efficient single-qubit rotations to allow extra freedom to prepare the ground-state accounting for Heisenberg interactions. This was found to be preferable to using a full Hamiltonian Variational ansatz~\cite{wecker2015progress}, which needed far more two-qubit gates to achieve a similar infidelity, due to requiring three $R^{zz}$ gates per bond per layer. We empirically observed added benefit from obtaining the ground state from an initially vortex-free state in the $J = 0.05$ case. In contrast, for the larger $J=0.2$ the further distance from the vortex-free sector showed less benefit from a vortex-free initialization. Instead, here we omitted both $U_1(\boldsymbol{\theta})$ and the projection operation $P_\text{VF}$, which were replaced with a single parameterized $R^{y}$ gate on each qubit at the start of the circuit.

\subsection{Ansatz optimization}
We optimize the circuit parameters using the L-BFGS optimizer to minimize the infidelity cost function 
\begin{align}
    C_\text{GS}(\bm \varphi, \bm \theta) & = 1-F \nonumber\\
    & = 1 - \frac{1}{\mathcal{N}}\big|\langle \psi_\text{GS}| U_2(\bm \varphi) P_\text{VF}U_1(\bm \theta)|\bm 0\rangle \big|^2,
\end{align}
obtained from GPU-accelerated statevector simulations, where $\ket{\psi_\text{GS}}$ is the ground state from an exact calculation and $\mathcal{N} = \bra{\bm 0} U_1^\dag(\bm \theta) P_\text{VF} U_1(\bm \theta)\ket{\bm 0}$ is a normalizing factor arising from the projection, and with a similar expression for the circuits used in Kitaev-Heisenberg state preparation. GPUs can significantly speedup simulating quantum circuits on large qubit numbers, however there is an extra time overhead due to data transfer between the CPU and GPU. For a conventional simulation of a deep quantum circuit this is typically negligible, however the implementation of the non-unitary projection amplifies this cost. 
To avoid constructing the full matrix of $P_\text{VF}$, the projection is simulated by iterating over each plaquette and removing the corresponding vortex, using statevector addition. While this only requires manipulated statevectors, avoiding the expensive construction of the full operator $P_\text{VF}$, the overhead due to CPU-GPU data transfers required for each plaquette projection significantly slows down the calculation. 

\begin{figure}[!htb]
    \centering
    \includegraphics[width=0.9\linewidth]{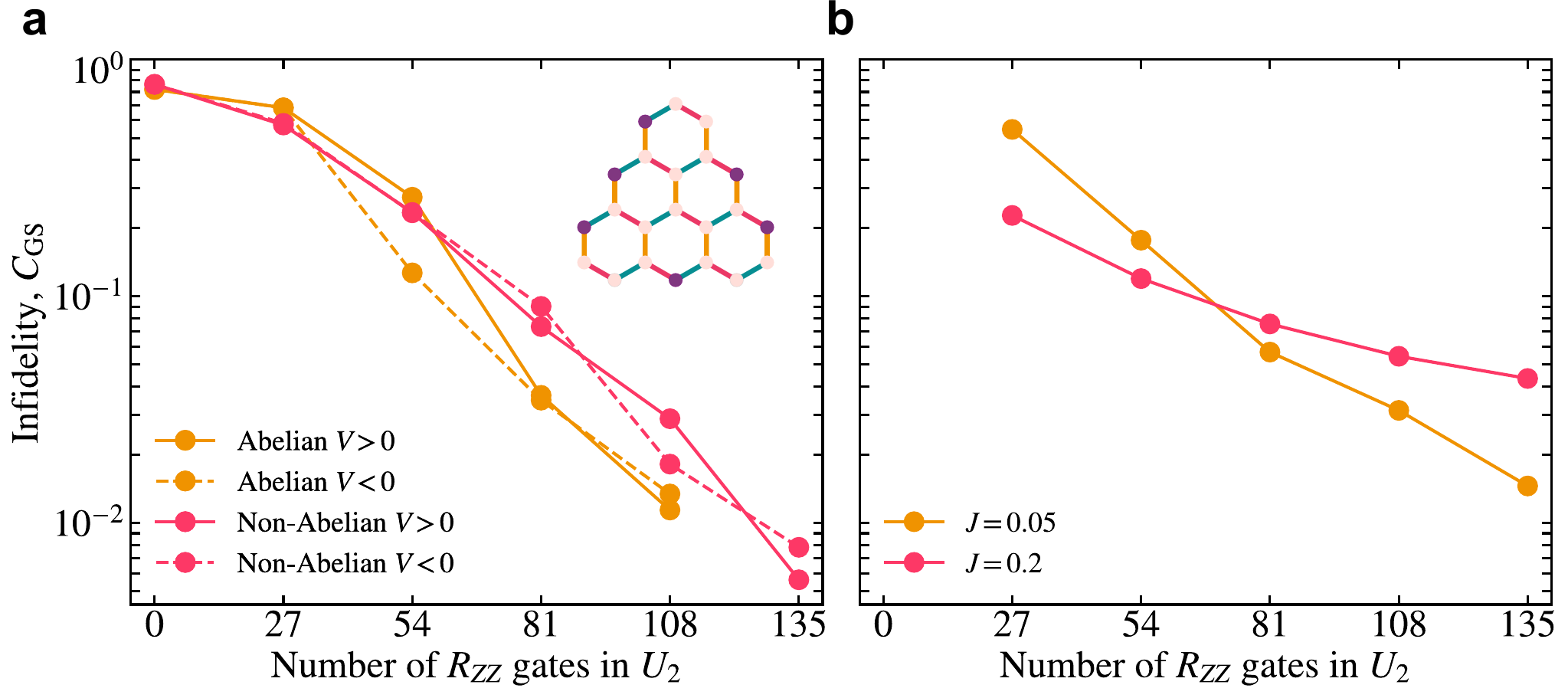}
    \caption{Ground state optimization.
    \textbf{a,} Ground state preparation error $C_\text{GS}$ for the pure Kitaev Hamiltonian ($J=0$) as the variational circuit depth is increased. 
    We consider four ground states, in the Abelian ($6K^x=6K^y=K^z=-1$) and non-Abelian ($K^x=K^y=K^z=-1$) phases, and with effective fields $V=\pm 0.3$. (inset) Purple qubits on the boundary denote correction qubits $c_p$ used to correct $W_p=-1$ plaquette operator measurements.
    \textbf{b,} Ground state optimization for the non-integrable Kitaev-Heisenberg Hamiltonian, with $K^x=K^y=K^z=-1, \ V=0.3$, and $J=0.05, 0.2$.}
    \label{fig:GS_Opt}
\end{figure}

To partially mitigate this cost while taking advantage of the variational freedom before the projection, we first optimize the simpler parameterized state $P_\text{VF}U_1(\bm \theta)|\bm 0\rangle$, e.g. $U_2(\bm \varphi) = I$. We found this optimization tractable as only 32 variational parameters need optimizing, significantly reducing the number of cost function calls requiring the expensive simulated projection operation. Despite the simplicity of the initial circuit dynamics $P_\text{VF}U_1(\bm \theta)|\bm 0\rangle$, we still find a respectable reduction in the cost, down to $C_\text{GS}=0.864$, as shown by the first datapoints on Fig.~\ref{fig:GS_Opt}. Then, with the converged $\bm \theta$ values, these are fixed and the $\bm \varphi$ are then optimized. As $P_\text{VF}U_1(\bm \theta)|\bm 0\rangle$ is fixed through the second stage of the optimization, this initial state is computed once before the optimization begins and saved, then allowing the cost function to be computed without the multiple CPU-GPU data transfers discussed above.
The optimization of $\bm \theta$ and $\bm \varphi$ separately allowed the variational freedom before the projection to be exploited somewhat, but not fully. In testing on smaller cluster sizes, without the constraints due to GPU simulation, $\bm \theta$ and $\bm \varphi$ were both optimized simultaneously and an even greater benefit was found, and in general this is the strategy we would advocate.

\section{Time Evolution Circuits}  \label{time evolution appendix}

\subsection{Exact Approach}
We derive an approximate time-evolution operator to generate the three-body interactions in $H_V$ using only hardware-native two-body unitary dynamics based on the terms in $H_K$. We begin by considering unitaries $U_\alpha(t_\alpha) = \exp(-i H_\alpha t_\alpha)$ with generators $H_\alpha = K^\alpha\sum_{\alpha\ \text{bonds}} \sigma^\alpha_i \sigma^\alpha_j$. We find an approximate propagator can be derived from the ansatz
\begin{equation} U(\tau) = U_z(\tau-t_z) U_y(\tau-t_y) U_x(\tau-t_x) U_z(t_z) U_y(t_y) U_x(t_x) \end{equation}
We equate $U(\tau) = e^{A}$ using the BCH expansion and find 
\begin{align} A =&  -i\tau (H_x + H_y + H_z) \nonumber\\
& - \frac{A_{zy}[H_z,H_y] + A_{zx}[H_z,H_x] + A_{yx}[H_y,H_x]}{2} + \ldots\end{align}
where $A_{\alpha\beta} = \tau^2 - 2t_\alpha(\tau-t_\beta).$ We then derive parameters $t_x, t_y,$ and $t_z$ to implement the propagator by setting $A = -i(H_K + H_V)\tau + \mathcal{O}(\tau^3,t\tau^2,t^2\tau,t^3)$, where the notation $t^m$ refers to any product of $m$ terms $t_x,t_y,t_z$. We find two sets of valid parameters
\begin{align} \label{final times} t_y^\pm & = \frac{\tau}{2}\left(1 \pm \sqrt{1-2 \frac{(1 + V/\tau K^zK^y)(1 + V/\tau K^yK^x)}{(1 - V/\tau K^zK^x)}}\right)  \nonumber\\
t_x^\pm & = \tau\left(1 - \frac{1 + V/\tau K^yK^x}{2t_y^\pm/\tau}\right)\nonumber\\
t_z^\pm & = t_y^\pm \frac{1 - V/\tau K^zK^x}{1 + V/\tau K^yK^x}
\end{align}
Solutions exist near the origin for $V > 0$ and for most negative $V$, where the term under the square root in $t_y^\pm$ is real.

The relations above assumed a unitary ordering $U_z U_y U_x$, in which $H_x$ dynamics are applied to the state before $H_y$ and $H_z$.  Similar relations can be derived for different unitary orderings. The resulting solutions are valid for distinct intervals of $V$ and achieve different accuracies in numerical simulations, which we attribute to higher order terms in the BCH expansions.  We found the ordering $U_z U_y U_x$ was most accurate for $V =0.3$ while $U_z U_x U_y$ was most accurate for $V =-0.3$, and we use these in our results.

To obtain the total propagator, including boundary field and Heisenberg interactions, we use Trotter approximations with additional unitaries to augment $U(\tau)$. To capture the effect of non-zero boundary terms $h$, we sandwich $U(\tau)$ with 
\begin{equation}
U_{h}(\tau) = \prod_{\text{{boundary}}} e^{-i h \tau \sigma^\alpha /2 },
\end{equation}
yielding a second-order approximation. These field terms incur a negligible expense in the quantum circuits, as the single-qubit gates needed for $U_h(\tau)$ have small error relative to two-qubit gates used in the other evolution operators.  

For each bond in $H_K$, the Heisenberg interactions alter the Kitaev bond strength to $K^\alpha \to K^\alpha + J$ while also introducing two new interactions $J\sigma^\beta_i\sigma^\beta_j$ and $J\sigma^\gamma_i\sigma^\gamma_j$ along distinct directions $\alpha\neq\beta\neq\gamma$. To implement the resulting dynamics in the circuit, we use a unitary operator $\tilde U(\tau)$ analogous to Eq.~(\ref{time_evol_ansatz}) except with modified interaction strengths $K^\alpha \to K^\alpha + J$.  We then append this with
\begin{equation}
U_{J}(\tau) = \prod_{\alpha\ \text{bonds}} e^{-i \tau J \sigma_{j}^\beta \sigma_{k}^\beta} e^{-i \tau J \sigma_{j}^\gamma \sigma_{k}^\gamma}
\end{equation}
 where $\alpha \neq \beta \neq \gamma \in \{ x, y, z \}$ in a first-order approximation.  In total this gives the approximate propagator
 \begin{equation}
U_\text{tot}(\tau) = U_{h}(\tau) \tilde U(\tau) U_{h}(\tau) U_{J} (\tau)
\end{equation}

The Trotterized expansion of $U_{h} (\tau)$ maintains the Trotter error at $ O(\tau^3)$, similar to the error in the analytic derivation of $U(\tau)$. The $U_{J}(\tau)$ incurs an error $\mathcal{O}(\tau^2)$, but with a coefficient $J\approx \tau$, hence the coefficients of the associated error terms are a similar size to the error coefficients in $U(\tau)$. Fig.~\ref{various_dts} demonstrates this ansatz converges to exact results as the timestep $\tau$ is decreased, as expected. 

\subsection{Variational Approach}
We now consider a scalable approach to numerically optimizing our ansatz, to further improve its accuracy for implementations on the quantum computer.  We consider optimization on a four-spin T-junction [Fig.~\ref{fig:optimized parameters} (inset)], to obtain optimized ansatz parameters that can be applied to the whole lattice, resulting in modest improvements relative to the previous analytic results.   For this optimization we consider the ansatz
\begin{equation}
U(\tau) = U_z(\tau - t_z') U_y(\tau - t_y') U_x(\tau - t_x') U_z(t_z) U_y(t_y) U_x(t_x).
\label{time_evol_unitary}
\end{equation}
defined on a T-junction, then optimize the $t_i,t_i'$ to minimize the 2-norm error with respect to an exact propagator containing two- and three-body terms in $H_K$ and $H_V$ acting only on the T-junction. The optimization yields $t_i' \neq t_i$ for $i \in \{x, y, z\}$, unlike the analytic solution. We find it is possible to achieve unitaries that are numerically equivalent to the exact T-junction unitary, with normalized 2-norm errors $\approx 10^{-11}$ (normalized by the 2-norm of the identity matrix with the same dimension). We rationalize the observed perfect agreement by noting that the BCH expansion contains only six terms on the T-junction, while the ansatz (Eq.~(\ref{time_evol_unitary})) contains the same number of free parameters. The parameter optimization therefore can be understood as tuning the coefficients of the six relevant bond and effective field terms, slightly modifying the analytic $t_i,t_i'$ to exactly account for higher order corrections. 

We then apply the ansatz (Eq.~(\ref{time_evol_unitary})) with the optimized $t_i,t_i'$ on the whole lattice, in place of the previous analytic expression, and quantitatively assess its performance. We again compute the normalized 2-norm error  between the optimized unitaries and the exact time evolution operator for systems of increasing size. Fig.~\ref{fig:optimized parameters} shows this comparison for both two- and three-plaquette systems across a range of effective field strengths $V$. For the 6-plaquette simulations, we are not able to compute the 2-norm error, due to the large size of the unitary operators.  We instead show in Fig.~\ref{fig:optimized parameters}\textbf{b} that the optimized unitary yields a modest but consistent enhancement in the correlation-function accuracy, relative to the analytical parameters. The improved agreement is consistent with accounting for some higher-order corrections in $H_K$ and $H_V$, as expected from the T-junction analysis, providing an enhanced ansatz for use on the quantum computer.

\section{Effective Temperature}\label{temperature appendix}

We assign an effective temperature to our Kitaev honeycomb simulations by comparing the measured state energies to the theoretical spectrum. The model is solved by expressing the Kitaev Hamiltonian in an expanded space, where each spin is represented as two complex fermions or four Majorana fermion operators, following Kitaev's original analysis \cite{kitaev2006anyons}. Only the \emph{physical} many-body states are retained, satisfying the global fermion-parity constraint
\begin{equation}
D = \det A (-1)^\theta\!\!\!\!\prod_{\langle j,k\rangle_\alpha}\!\!\!\!u_{jk}
\prod_{\ell=1}^{n}(1-2\,\psi_\ell^{\dagger}\psi_\ell)=+1,
\end{equation}
as derived in Ref.~\cite{pedrocchi2011physical} for appropriate boundary conditions. Here $\psi_\ell$ are the complex fermion modes obtained from the diagonalization of the Majorana Hamiltonian $H_\mathbf{F}$ (that includes $c$ type \& dangling $b^\alpha$ type majorana fermions), $u_{jk}=\pm1$ are the $\mathbb{Z}_2$ bond variables defining the background flux configuration, $(-1)^\theta$ is a lattice-geometry dependent factor that arises from anticommutation relations between the Majorana fermion operators, and $A$ the transformation matrix that diagonalizes the complex fermions. 

Using the free-fermion description~\cite{kitaev2006anyons}, we enumerate all $2^6$ flux configurations of the six plaquettes. Each configuration $\mathbf{F}$ defines a Hamiltonian $H_{\mathbf{F}}$ with couplings $K_{ij}\!\to\!\nu_{ij}K_{ij}$ and $V_{ik}\!\to\!\nu_{ij}\nu_{jk}V_{ik}$, where $\nu_{ij}(\mathbf{F})=\pm1$ are bond signs fixed by $\mathbf{F}$. For each configuration, $E_{\mathbf F, m}$ is the many-body eigenenergy of $m^{\text{th}}$ fermionic eigenstate within flux configuration $\mathbf{F}$ is computed consistent with the physical parity $D=+1$. From this complete spectrum, the canonical averages 
\begin{equation}
\langle H\rangle(\beta)=
\frac{\sum_{\mathbf F, m} e^{-\beta E_{\mathbf F,m}}E_{\mathbf F, m}}
     {\sum_{\mathbf F, m} e^{-\beta E_{\mathbf F, m}}},
\end{equation}
By equating the post-selected energy $E_{\mathrm{meas}}=-16.68$ with the thermal curve $\langle H\rangle(\beta)$, we extract $T_{\mathrm{eff}}\!\approx\!0.2|K_z|$ in our reduced units as shown in Fig.~\ref{fig:energy}(b). 

\begin{table*}[!htb]
\centering
\renewcommand{\arraystretch}{1.2}
\begin{tabular}{llcccc}
\hline
\diagbox{Contribution}{Time} &  & $t=0$ & $t=0.1$ & $t=0.3$ & $t=0.5$ \\
\hline
$\delta_{\mathrm{prep}}$ & Ground-state ansatz infidelity & 0.003 & 0.004 & 0.009 & 0.010 \\
$\delta_{\mathrm{Trotter}}$ & Trotter error, $\tau=0.1$ & 0.000 & 0.001 & 0.004 & 0.006 \\
$\delta_{\mathrm{compile}}$ & Three-body $\to$ two-body circuit & 0.000 & 0.017 & 0.050 & 0.084 \\
$\delta_{\mathrm{device}}^{\mathrm{raw}}$ & Hardware residual $-$ noiseless & 0.054 & 0.040 & 0.094 & 0.115 \\
$\delta_{\mathrm{device}}^{\mathrm{ps}}$ & Post-selected $-$ noiseless & 0.073 & 0.044 & 0.060 & 0.058 \\
$\delta_{\mathrm{total}}^{\mathrm{ps}}$ & Post-selected $-$ exact & 0.069 & 0.063 & 0.069 & 0.146 \\
\hline
$\sigma^{\mathrm{raw}}$ & Shot noise, $1\sigma$, all shots & 0.057 & 0.044 & 0.044 & 0.044 \\
$\sigma^{\mathrm{ps}}$ & Shot noise, $1\sigma$, post-selected & 0.066 & 0.053 & 0.059 & 0.069 \\
\hline
\end{tabular}
\caption{Error budget for the dynamical correlators $\langle Z_i(t)Z_{\mathrm C}\rangle$,
$i=\mathrm{R},\mathrm{C},\mathrm{L}$, in the non-Abelian phase [Fig.~\ref{fig:fig3}\textbf{b};
$K^{\alpha}=-1$, $V=0.3$, $h=0.1$, $J=0$]. Each entry is the modulus $\delta$ of the
complex deviation introduced by one approximation, quoted as the largest value over the
three correlators. See Appendix \ref{error budget} for details. }
\label{table: error_budget}
\end{table*}

\section{Exact Results and Error Budget} \label{error budget}

The exact results in Fig.~\ref{fig:energy} were obtained by computing the expectation values from the exact numerical ground state. To obtain the exact curves in Fig.~\ref{fig:fig3} and Fig.~\ref{fig:Fig4}, we first numerically obtain the exact ground state for each Hamiltonian, and then perform the time evolution using a 4th-order Trotter decomposition. We used a time step $dt = 0.1$ in Fig.~\ref{fig:fig3} and $dt = 0.075$ in Fig.~\ref{fig:Fig4}. We verified that these time steps yield converged results at the relevant time and correlation scales.

In Table~\ref{table: error_budget}, we decompose the various sources of error $\delta$ in the correlator magnitudes for a sample state, defining $\delta = \max_{i \in \{L,C,R\}} |\langle Z_i(t)Z_{\mathrm C}\rangle - \langle Z_i(t)Z_{\mathrm C}\rangle_\text{reference}|$ as the maximum error among correlators along the edge at a given time $t$. The reference is the exact numerical ground state evolved with a fourth-order Suzuki--Trotter product formula at $\tau=0.05$, converged to $O(10^{-7})$ which is negligible to table entries. The algorithmic rows are noiseless statevector simulations with one approximation enabled at a time and carry no statistical uncertainty; $\delta_{\mathrm{compile}}$ is quoted net of the T-junction re-optimization of the ansatz parameters. The device rows are the difference between the hardware data and a noiseless simulation of the same circuit, while the total represents the difference between the hardware and exact results.

The results show that the dominant source of algorithmic error is the time circuit compression to account for the three-body term using only two-body gates. The hardware results are comparable to shot-noise, with an improvement from post-selection appearing at longer time circuits.

\bibliography{main}

\section{Author Contributions}
P.C.L and A.B. conceived and supervised the project. G.B.H., B.X., and P.K. provided the theoretical framework. A.A. conducted the classical and quantum simulations. J.G. constructed the initial state circuits. K.K. constructed the time evolution circuits. V.M. conducted the effective temperature calculation and constructed the energy measurement circuits. All authors participated in the discussions and contributed to the manuscript.

\section{Competing Interests}
The authors declare no competing interests.

\onecolumngrid

\newpage

\onecolumngrid

\section{Supplemental Data}

\begin{table}[!htb]
\centering
\renewcommand{\arraystretch}{1.2} 
\begin{tabular}{lcccc}
\hline
\diagbox{Experiment}{\# of RZZ\\gates} 
& \shortstack{Ground\\State}
& \shortstack{Trotter\\Step}
& \shortstack{Vortex-Free\\Projection}
& \shortstack{Post-Selection\\Projection}
\\ \hline
Abelian                     & 108 & 54  & 36 & 36 \\ 
Non-Abelian $(V = 0.3)$     & 135 & 54  & 36 & 36 \\ 
Non-Abelian $(V = -0.3)$    & 108 & 54  & 36 & 36 \\ 
Heisenberg ($J = 0.05$)     & 81  & 108 & 36 & 0  \\ 
Heisenberg ($J = 0.2$)      & 81  & 108 & 0  & 0  \\ \hline
\end{tabular}
\caption{Number of RZZ gates used in each experiment.}
\label{table: gates}
\end{table}

\begin{figure}[!htb]
    \centering
    \includegraphics[width=0.6\linewidth]{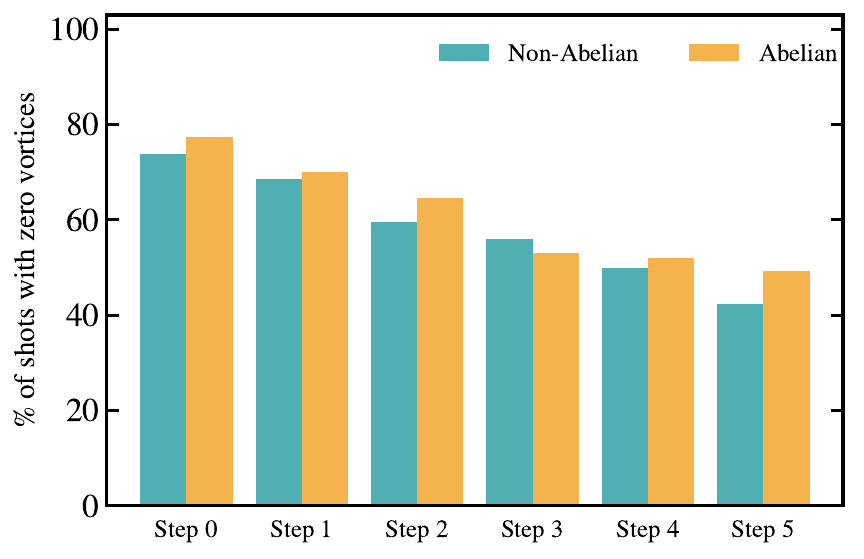}
    \caption{Plaquette vortex analysis. Percentage of shots with zero plaquette violations for every time step.}
    \label{fig:vortices}
\end{figure}

\begin{figure}[!htb]
    \centering
    \includegraphics[width=1\linewidth]{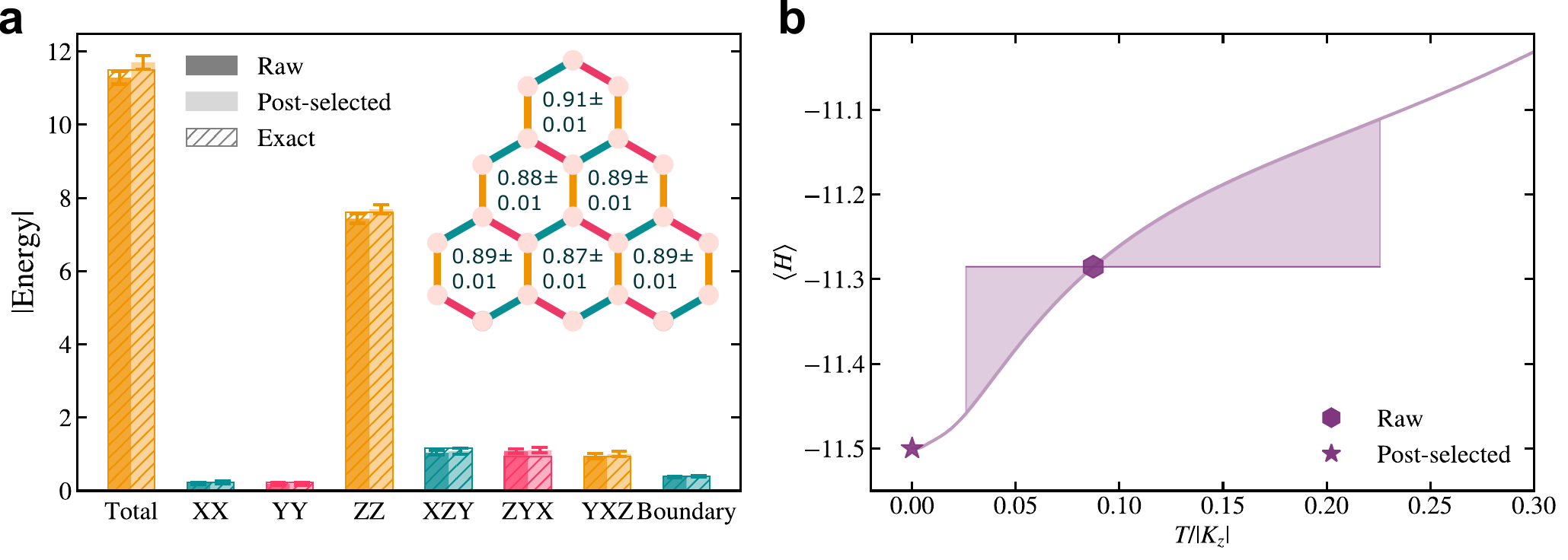}
    \caption{Abelian ground state preparation. \textbf{a,} Energy of the Abelian state with $6K^x=6K^y=K^z=-1,\ V= 0.3,\ J=0$ computed using H2-1 QPU. Error bars are the standard error. \textbf{b,} The effective temperatures corresponding to the raw and post-selected results. 
    }
    \label{fig:abelian-energy}
\end{figure}

\begin{figure}
    \centering
    \includegraphics[width=0.5\linewidth]{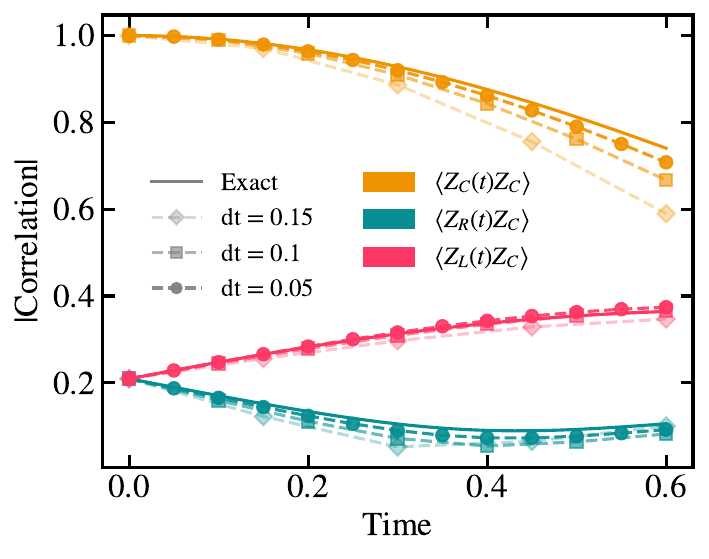}
    \caption{Convergence of the analytic time evolution circuit. Correlation functions converge to exact values as the timestep $\tau$ in $U_\text{tot}(\tau)$ decreases, for an example instance with $K^x = K^y = K^z = -1, V=0.3, h=0.1$, and $J=0$. The exact results are computed using Suzuki-Trotterization of the exact Hamiltonian with a timestep of 0.01. 
    }
    \label{various_dts}
\end{figure}

\begin{figure}[!htb]
    \centering
    \includegraphics[width=1\linewidth]{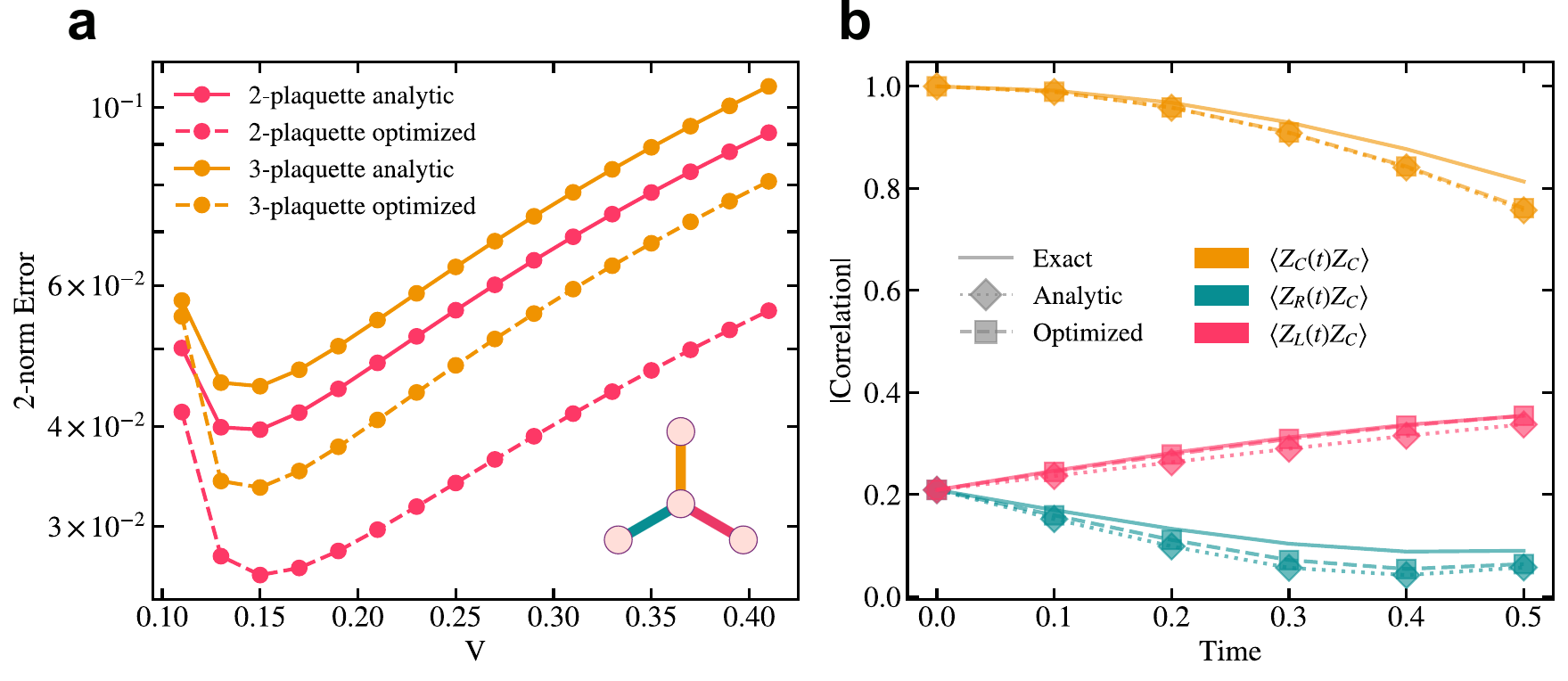}
    \caption{Comparing time evolution unitaries with optimized and analytical parameters. \textbf{a,} Optimizing unitary time-evolution parameters on a four-spin T-junction (inset) provides optimized parameters for use on the whole lattice. The accuracy of the resulting T-junction-optimized unitary on two- and three-plaquette systems is quantified by the difference 2-norm, normalized by the 2-norm of the identity matrix. The 2-norm error is smaller for the optimized unitary than for the unitary derived from analytic parameters.  Parameter are $( K^x,K^y,K^z,J,h)$ = $(-1,-1,-1,0,0)$ with various $V$.  \textbf{b,} Initializing the time-evolution unitaries with the optimized parameters improves the accuracy of the time-correlation function for the six-plaquette system for parameters $ K^x = K^y= K^z = -1,  J = 0, h =0.2$ , and $ V =0.3$. }
    \label{fig:optimized parameters}
\end{figure}

\begin{figure}[!htb]
    \centering
    \includegraphics[width=\linewidth]{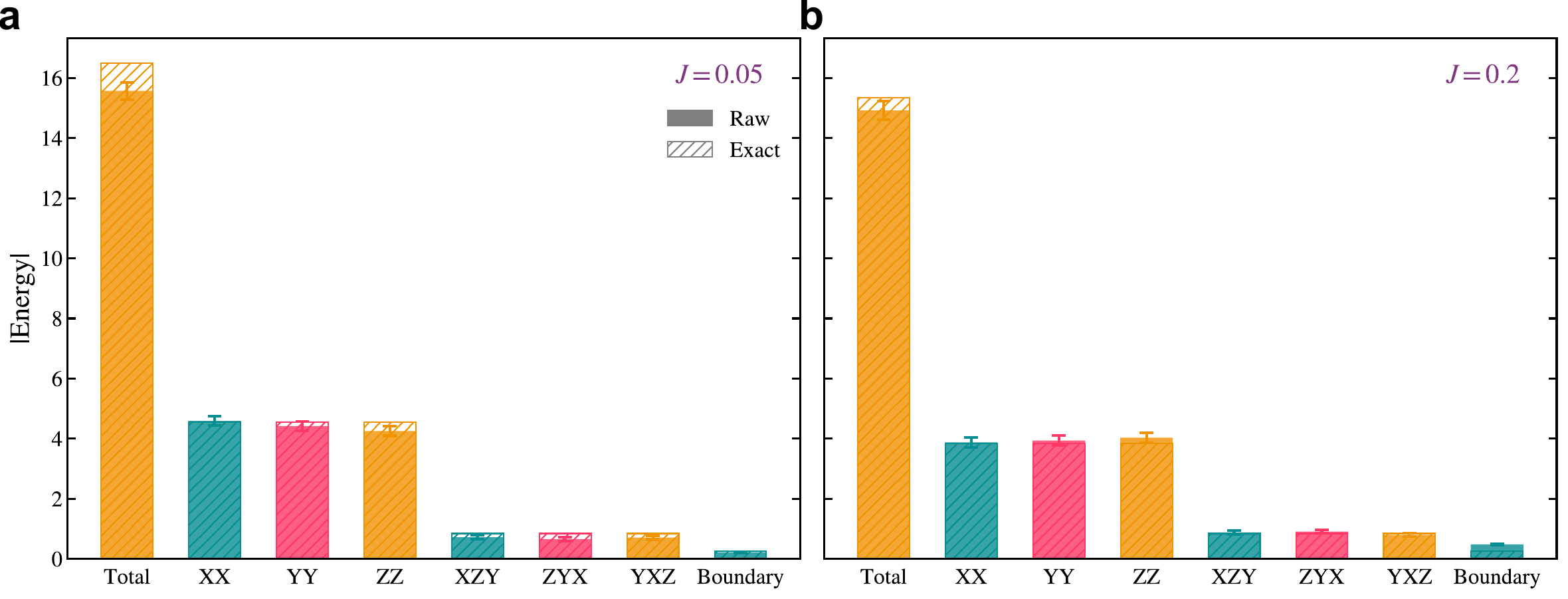}
    \caption{Heisenberg ground state preparation. \textbf{a,} Heisenberg ground state energy with $J=0.05$, and $K^x=K^y=K^z=-1,\ V= 0.3,\ h=0.1$. \textbf{b,} same parameters except for $J=0.2$.}
    \label{fig:heis_energies}
\end{figure}

\end{document}